\documentclass[longauth]{aa}
\usepackage{graphics,epsfig,txfonts,color}
\usepackage[dvips]{rotating}
\usepackage{natbib}
\bibpunct{(}{)}{;}{a}{}{,}

\newcommand{\ltsima}{$\buildrel < \over \sim$}
\newcommand{\lsim}{\lower.5ex\hbox{\ltsima}}
\newcommand{\gtsima}{$\buildrel > \over \sim$}
\newcommand{\gsim}{\lower.5ex\hbox{\gtsima}}

\newcommand{\hess}{H.E.S.S.}

\newcommand{\chandra}{\emph{Chandra}}

\newcommand{\fermi}{\emph{Fermi}}

\newcommand{\terzan}{Terzan\,5}
\newcommand{\htfive}{HESS\,J1747$-$248}

\begin{document}
\title{Search for Very-high-energy $\gamma$-ray emission from Galactic globular clusters with \hess}
\authorrunning{H.E.S.S. Collaboration}
\titlerunning{Search for $\gamma$-ray emission from Galactic GCs with \hess}
\author{H.E.S.S. Collaboration
\and A.~Abramowski \inst{1}
\and F.~Acero \inst{2}
\and F.~Aharonian \inst{3,4,5}
\and A.G.~Akhperjanian \inst{6,5}
\and G.~Anton \inst{7}
\and S.~Balenderan \inst{8}
\and A.~Balzer \inst{9,10}
\and A.~Barnacka \inst{11,12}
% ...
\email{Christopher.van.Eldik@physik.uni-erlangen.de}}
\institute{
Universit\"at Hamburg, Institut f\"ur Experimentalphysik, Luruper Chaussee 149, D 22761 Hamburg, Germany \and
Laboratoire Univers et Particules de Montpellier, Universit\'e Montpellier 2, CNRS/IN2P3,  CC 72, Place Eug\`ene Bataillon, F-34095 Montpellier Cedex 5, Fran
ce \and
Max-Planck-Institut f\"ur Kernphysik, P.O. Box 103980, D 69029 Heidelberg, Germany \and
Dublin Institute for Advanced Studies, 31 Fitzwilliam Place, Dublin 2, Ireland \and
National Academy of Sciences of the Republic of Armenia, Yerevan  \and
Yerevan Physics Institute, 2 Alikhanian Brothers St., 375036 Yerevan, Armenia \and
Universit\"at Erlangen-N\"urnberg, Physikalisches Institut, Erwin-Rommel-Str. 1, D 91058 Erlangen, Germany \and
University of Durham, Department of Physics, South Road, Durham DH1 3LE, U.K. \and
DESY, D-15735 Zeuthen, Germany \and
Institut f\"ur Physik und Astronomie, Universit\"at Potsdam,  Karl-Liebknecht-Strasse 24/25, D 14476 Potsdam, Germany \and
Nicolaus Copernicus Astronomical Center, ul. Bartycka 18, 00-716 Warsaw, Poland \and
CEA Saclay, DSM/Irfu, F-91191 Gif-Sur-Yvette Cedex, France
% ...
}

\date{Received / Accepted }
 
\abstract
{
Globular clusters (GCs) are established emitters of high-energy (HE, 100\,MeV$<$E$<$100\,GeV) $\gamma$-ray radiation which could originate from the cumulative emission of the numerous millisecond pulsars (msPSRs) in the clusters' cores or from inverse Compton (IC) scattering of relativistic leptons accelerated in the GC environment. 
These stellar clusters could also constitute a new class of sources in the very-high-energy (VHE, E$>$100\,GeV) $\gamma$-ray regime, judging from the recent detection of a signal from the direction of \terzan\ with the \hess\ telescope array. 
} 
{
To search for VHE $\gamma$-ray sources associated with other GCs, and to put constraints on leptonic emission models, we systematically analyzed the observations towards 15 GCs taken with the \hess\ array of imaging atmospheric Cherenkov telescopes. 
}
{
We searched for point-like and extended VHE $\gamma$-ray emission from each GC in our sample and also performed a stacking analysis combining the data from all GCs to investigate the hypothesis of a population of faint emitters. 
Assuming IC emission as the origin of the VHE $\gamma$-ray signal from the direction of \terzan , we calculated the expected $\gamma$-ray flux from each of the 15 GCs, based on their number of millisecond pulsars, their optical brightness and the energy density of background photon fields. 
}
{
We did not detect significant VHE $\gamma$-ray emission from any of the 15 GCs in either of the two analyses. 
Given the uncertainties related to the parameter determinations, the obtained flux upper limits allow to rule out the simple IC/msPSR scaling model for NGC\,6388 and NGC\,7078. 
The upper limits derived from the stacking analyses are factors between 2 and 50 below the flux predicted by the simple leptonic scaling model, depending on the assumed source extent and the dominant target photon fields. 
Therefore, \terzan\ still remains exceptional among all GCs, as the VHE $\gamma$-ray emission either arises from extra-ordinarily efficient leptonic processes, or from a recent catastrophic event, or is even unrelated to the GC itself. 
}
{}

\maketitle
\section{Introduction}
\label{sec-introduction}
Globular clusters (GCs) are very dense systems of stars, hosting the oldest and most evolved stellar populations of the Galaxy. 
The extremely high rate of close stellar encounters in the cores of GCs likely gives rise to large numbers of dynamically formed compact binary systems \citep{2006ApJ...646L.143P,2003ApJ...591L.131P,2010ApJ...714.1149H} which are believed to be the  progenitors of millisecond pulsars (msPSRs). 
Indeed, large msPSR populations have been detected in many Galactic GCs through radio observations  \citep[see][]{2005ASPC..328..147C,2008IAUS..246..291R}. 

A complementary approach to study the msPSR populations of GCs comes from high-energy (HE, 100\,MeV$<E<$100\,GeV) $\gamma$-ray observations. 
Because msPSRs can be strong sources of pulsed magnetospheric emission in this energy regime \citep[as shown by observations with \fermi /LAT,][]{2009Sci...325..848A}, $\gamma$-ray signals are expected from GCs either due to bright individual msPSRs or due to the cumulative emission from the whole population. 
Up to now, 14 Galactic GCs have been detected with \fermi /LAT \citep{2009Sci...325..845A,2010A&A...524A..75A,2010ApJ...712L..36K,2011ApJ...729...90T}, and the $\gamma$-ray spectra of most of these sources show steep cut-offs at a few GeV, very reminiscent of the spectra measured from individual msPSRs. 
Indeed, an individual $\gamma$-ray msPSR has recently been detected in the GC NGC\,6624 with \fermi /LAT \citep{2011Sci...334.1107F}. 

Through the observations of very-high-energy (VHE, $E>$100\,GeV) $\gamma$-rays even the whole msPSR population might be accessible. 
Models predicting emission in the VHE regime rely on inverse Compton (IC) up-scattering of soft photon fields by ultra-relativistic leptons accelerated in the pulsars' magnetospheres \citep{2009ApJ...696L..52V} or even re-accelerated in colliding pulsar wind nebula shocks \citep{2007MNRAS.377..920B}. 
Following \citet{2010ApJ...723.1219C}, under certain conditions even the HE emission observed by \fermi /LAT could be explained within a pure IC scenario. 
In all these models, depending on the assumed diffusion timescale, the magnetic field strength and the location of the GC, the relevant target photon fields could either be the GC starlight, the cosmic microwave background or the Galactic optical and infrared background light. 
In other words, if the leptons remain localized close to their sources, e.g. due to slow particle diffusion, the photon field of the GC member stars would be the dominant target for the IC process. 
In contrast, if the leptons propagate out of the core region of the GCs, other photon fields would become more important, because the energy density of the GC photon field rapidly decreases with the distance from the core. 
Thus, in this case the morphology and size of the VHE $\gamma$-ray source would be decoupled from the distribution of msPSRs as well as from the extent of the GC stellar population. 

Recently, VHE $\gamma$-ray emission has been detected from the direction of the GC \terzan\ with the \hess\ telescope array \citep[\htfive , ][]{2011A&A...531L..18H}. 
Should this signal indeed be associated to \terzan\ this would establish GCs as a new class of sources in VHE $\gamma$-rays, since previous studies of GCs by \hess , MAGIC, VERITAS, and CANGAROO in this energy domain resulted only in upper limits \citep[see e.g.][]{2009A&A...499..273A,2011ApJ...735...12A,2009ApJ...702..266A,2009arXiv0907.4974M,2007ApJ...668..968K}. 
Even though the nature of this source is not settled yet, IC emission from leptons accelerated by the particularly large msPSR population of \terzan\ remains a possibility. 
In this case also diffuse X-ray emission could be expected arising from synchrotron radiation (SR) from the same population of leptons \citep{2008AIPC.1085..277V}. 
Interestingly, extended and diffuse non-thermal X-ray emission was detected from \terzan\ with \chandra\ by \citet{2010A&A...513A..66E}. 
However, the apparent offset between the X-ray and VHE $\gamma$-ray signals challenges simple (IC/SR) models. 
A search for diffuse X-ray emission on similar scales from other HE-detected GCs resulted only in upper limits \citep{2012A&A...540A..17E}. 

As an alternative to the leptonic msPSR scenarios, a hadronic model was proposed for \htfive\ by \citet{2011A&A...533L...5D}. 
Here, it is suggested that the VHE $\gamma$-ray emission arises from the decay of neutral pions that were produced in inelastic collisions of relativistic protons with interstellar matter. 
These protons could have been accelerated in a remnant of a short $\gamma$-ray burst resulting from a merger of binary neutron stars. 
In the cores of GCs the probability for such a rare event is expected to be enhanced compared to Galactic average, due to their large binary fraction. 
With a rate for such rare catastrophic events of only $\sim$10$^{-4}$\,yr$^{-1}$ \citep{2007PhR...442..166N}, it is unlikely that two of them can be observed at the same time and, thus, \terzan\ may stand out as an exception among all GCs. 

At this point it is thus unclear if or to what extent the X-ray, HE and VHE $\gamma$-ray signals detected from GCs are related and what the underlying radiation mechanisms are. 
Further studies, particularly in the X-ray and VHE $\gamma$-ray regimes are the most promising to shed new light on the high-energy processes in these systems. 
In this work we present a systematic search for VHE $\gamma$-ray emission from all Galactic GCs that are covered by the present observational data set of the \hess\ telescope array. 
To increase the sensitivity for the detection of a population of individually faint emitters we also performed a stacking analysis, where we combined the number of photon candidate counts from the individual GCs to test for a statistically significant signal.
We then compare the obtained results to the VHE $\gamma$-ray signal detected from \terzan\ to investigate scenarios based on IC emission from leptons accelerated by msPSRs. 

\section{\hess\ observations}
\label{sec-observations}
\hess\ is an array of four imaging atmospheric Cherenkov telescopes located in the Khomas Highland of Namibia. 
Each telescope is equipped with a tessellated mirror with a surface area of 107\,m$^2$ and a camera comprising 960 photomultiplier tubes with a total field of view (FoV) of 5$^\circ$. 
For the reconstruction of atmospheric air showers the array works in coincidence mode requiring at least two telescopes to detect individual events. 
In this mode the array features a point-spread-function (PSF) with a 68\% containment radius of $\sim$0\fdg 07 and an energy resolution of 15\% on average. 
The stereoscopic reconstruction provides an excellent suppression of the cosmic-ray background reducing the number of background events after all cuts by a factor of up to 10$^4$ \citep{2005AIPC..745..611B}. 
With modern reconstruction techniques the \hess\ array has a point-source sensitivity of about 2$\times 10^{-13}$\,ph\,cm$^{-2}$\,s$^{-1}$ (E$>$200\,GeV) for 25 hours of observation, requiring a statistical significance of 5\,$\sigma$ \citep[see e.g.][]{2009APh....32..231D,2009APh....31..383O}. 
Here a zenith angle of 20$^\circ$ and an offset of of 0\fdg 7 is assumed. 

\begin{table}[tb]
\caption[]{The GC sample studied in this work}
\renewcommand{\tabcolsep}{3.5pt}
\begin{center}
\begin{tabular}{llllll}
\hline\hline\noalign{\smallskip}
\multicolumn{1}{l}{GC} &
\multicolumn{1}{l}{long.$^{(1)}$} &
\multicolumn{1}{l}{lat.$^{(1)}$} &
\multicolumn{1}{l}{zenith$^{(2)}$} &
\multicolumn{1}{l}{offset$^{(2)}$} &
\multicolumn{1}{l}{livetime$^{(3)}$} \\
\multicolumn{1}{l}{name} &
\multicolumn{1}{l}{($^{\circ}$)} &
\multicolumn{1}{l}{($^{\circ}$)} &
\multicolumn{1}{l}{($^{\circ}$)} &
\multicolumn{1}{l}{($^{\circ}$)} &
\multicolumn{1}{l}{(h)} \\
\noalign{\smallskip}\hline\noalign{\smallskip}
NGC\,104 (47\,Tuc)$^{(*)}$        &       305.89  &       $-$44.89        &       49.8    &       0.9     &       23.1    \\
NGC\,6388$^{(*)}$       &       345.56  &       $-$6.74 &       23.6    &       0.7     &       17.9    \\
NGC\,7078 (M\,15)       &       65.01   &       $-$27.31        &       37.5    &       0.7     &       12.3    \\
Terzan\,6 (HP\,5)       &       358.57  &       $-$2.16 &       24.6    &       1.8     &       15.2    \\
Terzan\,10&     4.49    &       $-$1.99 &       18.1    &       1.6     &       4.2             \\
NGC\,6715 (M\,54)      &       5.61    &       $-$14.09        &       18.3    &       0.7     &       11.8    \\
NGC\,362        &       301.53  &       $-$46.25        &       49.5    &       1.3     &       5.0             \\
Pal\,6  &       2.10    &       1.78    &       19.4    &       1.4     &       24.7    \\
NGC\,6256       &       347.79  &       3.31    &       20.5    &       1.3     &       5.3             \\
Djorg\,2        &       2.77    &       $-$2.50 &       15.9    &       1.7     &       4.6     \\
NGC\,6749       &       36.20   &       $-$2.21 &       33.7    &       1.5     &       8.2     \\
NGC\,6144       &       351.93  &       15.70   &       26.8    &       1.4     &       4.7             \\
NGC\,288        &       152.30  &       $-$89.38        &       11.8    &       1.4     &       46.7    \\
HP\,1 (BH\,229)          &       357.44  &       2.12    &       14.3    &       1.5     &       5.6     \\
Terzan\,9       &       3.61    &       $-$1.99 &       18.4    &       1.5     &       5.2     \\
\hline\noalign{\smallskip}
\end{tabular}
\label{tab-gc-sample}
\end{center}
$^{(1)}$Galactic coordinates used for the analyses, as given by H10;
$^{(2)}$Mean zenith angle and offset averaged over all contributing runs;
$^{(3)}$Acceptance corrected livetime of all \hess\ observation runs passing quality cuts;
$^{(*)}$GC is detected with \fermi /LAT.
\end{table}

\subsection{The globular cluster sample}
\label{sec-gc-sample}
We based our target selection on the GC catalog of \citet[][henceforth H10]{1996AJ....112.1487H} from which we also took all basic parameters such as position, distance ($d$), core radius ($r_\mathrm{c}$), and central luminosity density ($\rho_\mathrm{0}$). 
To select the sample of GCs and observational data that are suited for our study with \hess\ we applied the following \emph{a priori} cuts on the target and observation run list:

\begin{enumerate}
        \item GC $\left|\mathrm{Galactic~latitude}\right|$\,$\ge$\,1\fdg 0
        \item Run passes standard quality selection criteria and is pointed $<$2\fdg 0 offset from GC position
        \item GC has at least 20 available runs passing cut 2 
\end{enumerate}

The first cut is to conservatively exclude GCs towards the direction of the Galactic plane to avoid chance coincidences with unrelated sources and to prevent potential contamination from a population of faint unresolved sources and/or diffuse emission. 
Following the same approach as in the case of \terzan\ \citep[see][]{2011A&A...531L..18H} we find that the probability of an individual GC coinciding by chance with an unrelated Galactic source based on its latitude is less than 10$^{-3}$ for all GCs in this sample. 
The second cut excludes GCs that were only observed at very large offset angles from the camera center, because such observation runs would only provide relatively low effective exposure. 
With the third cut we make sure that there is a reasonably long exposure on these potentially faint sources. 
In addition we excluded \terzan\ from this study because those results were recently covered in a separate paper \citep[see][]{2011A&A...531L..18H}. 
Table~\ref{tab-gc-sample} lists the 15 GCs passing these cuts. 
We note that one of the GCs, M\,54, is actually not Galactic but most likely belongs to Sagittarius Dwarf Elliptical Galaxy \citep{1994Natur.370..194I}.

Most of the GC positions were either serendipitously covered in the \hess\ Galactic plane survey \citep[see][]{2006ApJ...636..777A,2012arXiv1204.5860G} or lie in the same FoV as other sources observed with \hess\ 
Therefore, the offsets of the GCs with respect to the camera center show a larger run-by-run scatter and in many cases larger mean values (see Tab.~\ref{tab-gc-sample}) compared to dedicated source observations. 
However, with a maximum allowed offset of 2\fdg 0 for individual runs (i.e. ``cut 2") the dataset is still in a range where the analysis is very well calibrated. 
Also, the \hess\ observations have been carried out during a time span of about six years where the gains of the photo multipliers and the optical reflectivity of the mirrors changed. 
To correct for these effects we calibrate the energy scale of the instrument using the response to single muons \citep[see][]{2006A&A...457..899A}. 

\begin{table*}[t]
\caption[]{Analysis results and model predictions}
\begin{center}
\begin{tabular}{llllllllll}
\hline\hline\noalign{\smallskip}
\multicolumn{1}{l}{GC} &
\multicolumn{1}{l}{E$_\mathrm{th}^{(1)}$} &
\multicolumn{1}{l}{${N_\mathrm{ON}}^{(2)}$} &
\multicolumn{1}{l}{${N_\mathrm{OFF}}^{(2)}$} &
\multicolumn{1}{l}{1/$\alpha^{(3)}$} &
\multicolumn{1}{l}{sig.$^{(4)}$} &
\multicolumn{1}{l}{$r^{(5)}$} &
\multicolumn{1}{l}{${F_\mathrm{UL}(E>\mathrm{E}_\mathrm{th})}^{(6)}$} &
\multicolumn{1}{l}{${F_\mathrm{UL} / F_\mathrm{IC;GC}}^{(7)}$} &
\multicolumn{1}{l}{${F_\mathrm{UL} / F_\mathrm{IC;IR,opt,CMB}}^{(7)}$} \\
\multicolumn{1}{l}{name} &
\multicolumn{1}{l}{(TeV)} &
\multicolumn{2}{c}{(counts)} &
\multicolumn{1}{l}{} &
\multicolumn{1}{l}{($\sigma$)} &
\multicolumn{1}{l}{($^\circ$)} &
\multicolumn{1}{c}{(ph\,cm$^{-2}$\,s$^{-1}$)} &
\multicolumn{2}{l}{} \\
\noalign{\smallskip}\hline\noalign{\smallskip}
\multicolumn{4}{l}{\emph{a) point-like source analysis}} \\
NGC\,104        &       0.72    &       72      &       941     &       18.2    &       2.6             &--& 1.9$\times 10^{-12}$    &       2.6$\times 10^{-1}$       &       2.1$\times 10^{1}$    \\
NGC\,6388       &       0.28    &       180     &       2365    &       14.9    &       1.6             &--& 1.5$\times 10^{-12}$    &       8.0$\times 10^{-2}$       &       1.6$\times 10^{0}$    \\
NGC\,7078       &       0.40    &       119     &       1988    &       15.0    &       $-$1.2          &--& 7.2$\times 10^{-13}$    &       1.9$\times 10^{-1}$       &       2.1$\times 10^{1}$    \\
Terzan\,6       &       0.28    &       202     &       8194    &       42.0    &       0.5             &--& 2.1$\times 10^{-12}$    &       7.3$\times 10^{-1}$       &       1.0$\times 10^{0}$    \\
Terzan\,10      &       0.23    &       76      &       2455    &       36.0    &       0.9             &--& 2.9$\times 10^{-12}$    &       4.3$\times 10^{-1}$       &       2.7$\times 10^{-1}$   \\
NGC\,6715       &       0.19    &       159     &       2361    &       15.2    &       0.3             &--& 9.3$\times 10^{-13}$    &       3.1$\times 10^{-1}$       &       1.3$\times 10^{2}$    \\
NGC\,362        &       0.59    &       18      &       533     &       33.0    &       0.4             &--& 2.4$\times 10^{-12}$    &       3.9$\times 10^{0}$        &       1.8$\times 10^{2}$   \\
Pal\,6          &       0.23    &       363     &       10810   &       31.4    &       1.0             &--& 1.2$\times 10^{-12}$    &       1.3$\times 10^{1}$        &       1.1$\times 10^{1}$   \\
NGC\,6256       &       0.23    &       64      &       1869    &       27.4    &       $-$0.5          &--& 3.2$\times 10^{-12}$    &       1.8$\times 10^{1}$        &       2.9$\times 10^{1}$   \\
Djorg\,2        &       0.28    &       56      &       2387    &       39.4    &       $-$0.6          &--& 8.4$\times 10^{-13}$    &       1.0$\times 10^{1}$        &       1.0$\times 10^{1}$   \\
NGC\,6749       &       0.19    &       84      &       2633    &       29.3    &       $-$0.6          &--& 1.4$\times 10^{-12}$    &       2.5$\times 10^{1}$        &       4.1$\times 10^{1}$   \\
NGC\,6144       &       0.23    &       63      &       2196    &       30.8    &       $-$1.0          &--& 1.4$\times 10^{-12}$    &       3.8$\times 10^{2}$        &       1.1$\times 10^{3}$   \\
NGC\,288        &       0.16    &       647     &       24148   &       38.5    &       0.8             &--& 5.3$\times 10^{-13}$    &       2.7$\times 10^{2}$        &       3.2$\times 10^{3}$   \\
HP\,1           &       0.23    &       67      &       2771    &       34.3    &       $-$1.6          &--& 1.5$\times 10^{-12}$    &       5.2$\times 10^{2}$        &       1.7$\times 10^{2}$   \\
Terzan\,9       &       0.33    &       89      &       2556    &       31.7    &       0.9             &--& 4.5$\times 10^{-12}$    &       2.6$\times 10^{4}$        &       9.0$\times 10^{2}$   \\
\hline\noalign{\smallskip}
\multicolumn{4}{l}{\emph{b) extended source analysis}}\\
NGC\,104        &       "        &   293 &   2016        &   7.4         &   1.2        & 0.22	 &	2.3$\times 10^{-12}$	 &	 2.3$\times 10^{-1}$	  &	  1.9$\times 10^{1}$	  \\
NGC\,6388       &       "        &   253 &   2818        &   12.9        &   2.2        & 0.11	 &	1.7$\times 10^{-12}$	 &	 9.2$\times 10^{-2}$	  &	  1.8$\times 10^{0}$	  \\
NGC\,7078       &       "        &   161 &   2386        &   14.0        &   $-$0.7     & 0.11	 &	1.1$\times 10^{-12}$	 &	 2.8$\times 10^{-1}$	  &	  3.1$\times 10^{1}$	  \\
Terzan\,6       &       "        &   304 &   9802        &   34.2        &   1.0        & 0.12	 &	2.4$\times 10^{-12}$	 &	 8.1$\times 10^{-1}$	  &	  1.2$\times 10^{0}$	  \\
Terzan\,10      &       "        &   218 &   4134        &   19.0        &   0.0        & 0.18	 &	3.6$\times 10^{-12}$	 &	 5.4$\times 10^{-1}$	  &	  3.4$\times 10^{-1}$	    \\
NGC\,6715       &       "    	 &   159 &   2361        &   15.2        &   0.3        & *      &      9.3$\times 10^{-13}$    &       3.1$\times 10^{-1}$       &       1.3$\times 10^{2}$    \\
NGC\,362        &       "        &   30  &   708         &   25.6        &   0.4        & 0.13	 &	2.5$\times 10^{-12}$	 &	 4.0$\times 10^{0}$	 &	 1.8$\times 10^{2}$	 \\
Pal\,6          &       "        &   1148&   17631       &   16.6        &   2.5        & 0.18 	 &	2.1$\times 10^{-12}$	 &	 2.4$\times 10^{1}$	 &	 1.9$\times 10^{1}$	 \\
NGC\,6256       &       "        &   131 &   2524        &   20.4        &   0.6        & 0.13	 &	3.9$\times 10^{-12}$	 &	 2.1$\times 10^{1}$	 &	 3.5$\times 10^{1}$	 \\
Djorg\,2        &       "        &   137 &   3753        &   24.8        &   $-$1.2     & 0.16	 &	9.7$\times 10^{-13}$	 &	 1.2$\times 10^{1}$	 &	 1.2$\times 10^{1}$	 \\
NGC\,6749       &       "        &   168 &   3544        &   20.7        &   $-$0.3     & 0.14	 &	2.1$\times 10^{-12}$	 &	 3.6$\times 10^{1}$	 &	 5.9$\times 10^{1}$	 \\
NGC\,6144       &       "        &   120 &   2913        &   23.9        &   $-$0.2     & 0.13	 &	2.5$\times 10^{-12}$	 &	 6.7$\times 10^{2}$	 &	 1.9$\times 10^{3}$	 \\
NGC\,288        &       "        &   1030&   30767       &   30.7        &   0.8        & 0.13	 &	6.1$\times 10^{-13}$	 &	 3.1$\times 10^{2}$	 &	 3.7$\times 10^{3}$	 \\
HP\,1           &       "        &   67  &   2771        &   34.3        &   $-$1.6     & *   	 &	1.5$\times 10^{-12}$	 &	 5.2$\times 10^{2}$	 &	 1.7$\times 10^{2}$	 \\
Terzan\,9       &       "        &   206 &   3909        &   18.8        &   $-$0.1     & 0.16	 &	4.1$\times 10^{-12}$	 &	 1.8$\times 10^{4}$	 &	 6.2$\times 10^{2}$	 \\
\hline\noalign{\smallskip}
\multicolumn{9}{l}{\emph{stacking analysis}}\\
\emph{a)}       &       0.23    &       2242    &       67826   &       31.2    &       1.6     &       --      &       3.3$\times 10^{-13}$    &       (5.4$^{+16}_{-1.7}$)$\times 10^{-2}$     &       (4.3$^{+11}_{-1.4}$)$\times 10^{-1}$     \\[3pt]
\emph{b)}       &       "       &       4425    &       92037   &       21.6    &       2.4     &       --      &       4.5$\times 10^{-13}$    &       (7.5$^{+23}_{-2.4}$)$\times 10^{-2}$     &       (5.9$^{+17}_{-2.0}$)$\times 10^{-1}$     \\[3pt]
\hline\noalign{\smallskip}
\end{tabular}
\label{tab-analysis-results}
\end{center}
$^{(1)}$Energy threshold of the analysis, defined as the location of the peak in the distribution of reconstructed photon energies;
$^{(2)}$Total number of on- and off-counts; 
$^{(3)}$Ratio between off- and on-exposure when applying the \emph{reflected background} technique;
$^{(4)}$Detection significance (pre-trial) following \citet{1983ApJ...272..317L};
$^{(5)}$Extraction radius used for the analysis assuming extended emission; a star(*) denotes the cases where the calculated intrinsic extent is negligible compared to the PSF (see sect.~\ref{sec-analysis-individual});
$^{(6)}$Photon flux upper limits \citep[99\% c.l., following][]{1998PhRvD..57.3873F} assuming a power-law spectrum with an index of $-$2.5;
$^{(7)}$Ratio between the upper limit and the expected IC flux using either the GC starlight (GC), or the sum of the Galactic infrared (IR), Galactic optical (opt) and the CMB background as target photon fields (see sect.~\ref{sec-discussion}). 
\end{table*}

\subsection{Analysis of individual GCs}
\label{sec-analysis-individual}
We applied the standard quality selection cuts on the \hess\ observational runs to reject data affected by bad weather conditions or poor instrument performance \citep[see][]{2006A&A...457..899A}. 
For the analysis we used the \emph{model} technique where the air showers are described by a semi-analytical model. 
The expected camera images are then compared to the observational data based on a maximum likelihood method \citep[see][]{2009APh....32..231D}. 
The \emph{model} analysis yields an improved sensitivity, particularly at lower energies, and a more efficient Gamma-Hadron separation compared to the standard Hillas approach. 
We used standard cuts for the \emph{model} analysis which include a 60 photo electron cut on the image size. 
With this configuration the \emph{model} analysis features a PSF with a 68\% containment radius of 0\fdg 07 on average. 
We cross-checked all results with an independent calibration using a Hillas analysis framework, which employs a machine-learning algorithm based on Boosted Decision Trees \citep{2009APh....31..383O}. 

\begin{figure}[tb]
        \begin{center}
                \resizebox{0.7\hsize}{!}{\includegraphics[clip=]{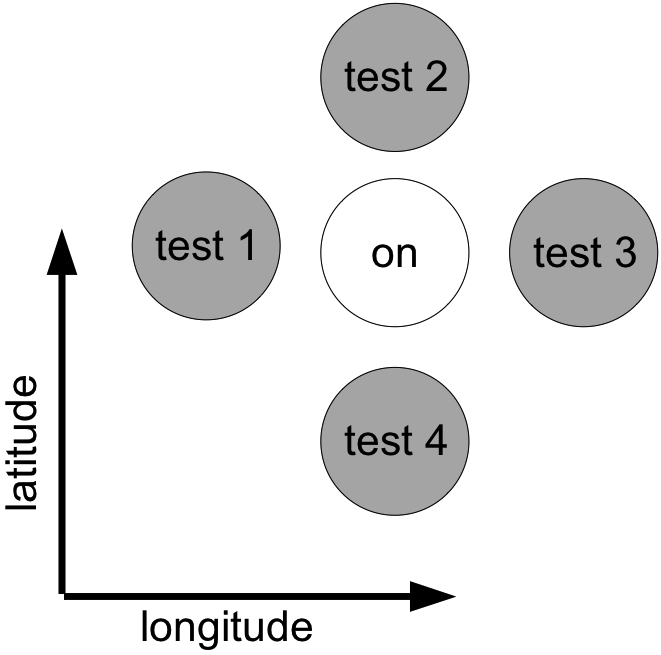}}
                \caption{Scheme showing the orientation of the four offset test regions with respect to the on-source region, aligned to the Galactic coordinate system. 
                The offsets are 0\fdg 4 in both directions and are defined such that region ``test\,4" is always closer to the Galactic plane than region ``test\,2". }
        \label{fig-offset-scheme}
        \end{center}
\end{figure}

To search for VHE $\gamma$-ray emission from each GC we performed two different kinds of analyses where the size of the emission region was assumed to be either a) point-like (0\fdg 1 integration radius) or b) extended while using the nominal GC position as the center of the extraction region (see Tab.~\ref{tab-gc-sample}). 
For case b) we determined the expected intrinsic size (with the effects of the PSF removed) of the source related to each GC based on the measured extent of the VHE $\gamma$-ray source \htfive\ detected from the direction of \terzan\ \citep{2011A&A...531L..18H}. 
The morphology of \htfive\ was fitted with an asymmetrical 2-dimensional Gaussian with intrinsic major and minor axes of 0\fdg 16 and 0\fdg 03, respectively. 
Assuming that the physical size of the VHE $\gamma$-ray source is the same among all GCs, we scaled the angular size of the major axis of \htfive\ with the relative distance of each individual GC (as given by H10) and \terzan\ \citep[$d_\mathrm{T5} = 5.9$\,kpc, ][]{2009Natur.462..483F}. 
We note here that \htfive\ is not only extended but also offset by 4\farcm 0$\pm$1\farcm 9 from the center of \terzan . 
Should such offsets be a general feature of VHE $\gamma$-ray emission from GCs this would also affect the regions for analysis b). 
However, the reason for this offset has not been attributed to any physical parameter of \terzan\ and particularly its direction for other GCs is therefore not easily predictable. 
Keeping the regions centered on the nominal GC positions but increasing the radius to also account for a potential offset could increase the fraction of the emission covered by the extraction region.
However, this would also heavily increase the number of background counts, thereby decreasing the sensitivity. 
Due to this fact, and since the offset seen from \terzan\ is much less significant than the intrinsic extent of the signal, we decided to not include the offset in the determination of the expected source size. 
To define the final extraction region for analysis b) we assumed a 2-dimensional gaussian morphology with the width of the expected intrinsic size, folded with the \hess\ PSF. 
The radius of the extraction region was then the 68\% containment radius of the folded distribution. 
These radii are listed in Tab.~\ref{tab-analysis-results}. 
Note that for the two GCs NGC\,6715 and HP\,1 the calculated intrinsic size was negligible ($<1\%$ of the size of the PSF) and we therefore considered the point-like analysis results also for case b). 

For the analyses we used the \emph{reflected background} technique \citep[see][]{2007A&A...466.1219B}, which ensures that the regions used for signal (ON) and background (OFF) extraction have the same acceptance in the FoV of the camera. 
In Tab.~\ref{tab-analysis-results} we present the results for both analyses and for each individual GC. 
From none of the GCs, neither with analysis a) nor b), significant excess emission is seen above the estimated background, as the individual significances for a detection, even pre-trials, are well below the threshold of 5\,$\sigma$. 
The mean (rms) values of the significance distributions are 0.30 (1.0)\,$\sigma$ and 0.34 (1.1)\,$\sigma$ for analysis a) and b), respectively. 
Thus, the significance distributions are reasonably compatible with only background fluctuations within statistical uncertainties. 
We derived upper limits for the photon flux above the energy threshold \citep[99\% confidence level, following][]{1998PhRvD..57.3873F} assuming a power-law spectrum for the photon flux
\begin{equation}
	F_\mathrm{ph} \propto \left(\frac{E}{\mathrm{TeV}}\right)^\Gamma
\end{equation}
with an index of $\Gamma = -2.5$, which we chose to allow for a comparison with the results obtained for \terzan\ \citep[$\Gamma_\mathrm{T5} = -2.5 \pm 0.3_\mathrm{stat} \pm 0.2_\mathrm{sys}$,][]{2011A&A...531L..18H}. 
Varying the assumed spectral index between $-$2.0 and $-$3.0 changes the upper limit by only 10\% to 15\%.

\begin{table}[tb]
\caption[]{Results for on-source and offset regions}
\begin{center}
\begin{tabular}{l|lll|lll}
\hline\hline\noalign{\smallskip}
\multicolumn{1}{l}{} &
\multicolumn{3}{|l}{\emph{point-like}} &
\multicolumn{3}{|l}{\emph{extended}} \\
\multicolumn{1}{l}{region} &
\multicolumn{1}{|l}{mean} &
\multicolumn{1}{l}{rms} &
\multicolumn{1}{l}{sig. stack} &
\multicolumn{1}{|l}{mean} &
\multicolumn{1}{l}{rms} &
\multicolumn{1}{l}{sig. stack} \\
\multicolumn{1}{l}{} &
\multicolumn{3}{|c}{($\sigma$)} &
\multicolumn{3}{|c}{($\sigma$)} \\
\noalign{\smallskip}\hline\noalign{\smallskip}
on      &       0.30    &       1.0     &       1.6     &       0.34    &       1.1     &       2.4     \\
test\,1 &       $-$0.16 &       1.4     &       $-$0.13 &       $-$0.12 &       1.2     &       $-$0.26 \\
test\,2 &       $-$0.39 &       1.0     &       $-$0.93 &       $-$0.15 &       1.1     &       0.55    \\
test\,3 &       $-$0.35 &       1.1     &       $-$0.41    &       $-$0.06 &       1.2     &       0.8     \\
test\,4 &       $-$0.31 &       1.3     &       $-$0.86 &       $-$0.28 &       1.2     &       $-$1.0  \\
\hline\noalign{\smallskip}
\end{tabular}
\label{tab-offset-results}
\end{center}
Mean and rms of the detection significance distributions for the on-source and offset test regions. Also, the total significances obtained with the stacking analyses are quoted. 
\end{table}

As a cross check, we performed analyses with identical parameters for four test positions in the vicinity of each GC. 
These regions are offset by $\pm$0\fdg 4 in Galactic coordinates (see Fig.~\ref{fig-offset-scheme}). 
At this distance in no case a contamination from the supposed GC sources would be expected. 
The purpose of these test positions is to generate a sample of source-free regions with a similar distribution in Galactic coordinates as the GC positions while still using a dataset comparable to the on-source regions. 
Also, because the region test\,4 is always closer to the Galactic plane than the other regions, particularly test\,2, one can study the effects that the Galactic latitude may have on this particular population study. 
In Tab.~\ref{tab-offset-results} we show the parameters of the significance distributions for the on-source and the four test regions. 
As is evident from these results, we did not detect a significant signal from any of these regions. 
Even though there appears to be a slight tendency towards negative significances for the test regions, the absolute mean values are very small and nothing substantial can be concluded from that apparent shift. 
To estimate the background for the test regions we excluded the on-source area. 
Therefore, the slight shift towards negative significances for test\,1 to test\,4 cannot be explained by a faint signal related to the GCs, which could have otherwise led to an overestimation of the background. 

\subsection{Stacking analysis}
\label{sec-analysis-stacking}
Because no individual GC studied in this sample shows significant VHE $\gamma$-ray emission, we performed a stacking analysis to search for a population of faint emitters. 
To combine the analysis results and count-statistics from the observation runs of all GCs we applied the same method that is used to combine the results for analyses of individual sources: 
The total number of ON (OFF) counts is the sum over all ON (OFF) counts of the runs in the stack. 
We calculated the ratio between the ON and OFF exposure ($\alpha$) as the weighted mean of $\alpha$ of all observation runs with the weight being the number of OFF counts per run. 
The energy-dependent effective area of the stack is calculated as the average over all runs, weighted with the livetime. 

The total GC stack has an acceptance-corrected livetime of 195 hours of good quality data and features an energy threshold of 0.23\,TeV. 
We performed the stacking analysis for both the point-like and extended source analyses described in the previous section. 
The results are shown in Tab.~\ref{tab-analysis-results}. 
The total detection significances obtained with these stacking analyses are 1.6\,$\sigma$ and 2.4\,$\sigma$, respectively, which are also very well below the threshold of 5\,$\sigma$ required for a firm detection. 
Therefore, we calculated upper limits on the photon flux for the full stack, again assuming a power-law spectrum with an index of $\Gamma = -2.5$ (see Tab.~\ref{tab-analysis-results}). 

For comparison we also performed the stacking analyses on each of the four offset test regions as defined in the previous section. 
The stacked significances obtained for these regions are systematically lower than for the on-source positions (see Tab.~\ref{tab-offset-results}). 
However, the differences are not sufficiently significant to draw any firm conclusions at this point. 

\section{Discussion}
\label{sec-discussion}
Even though neither any individual GC nor the stack as a whole revealed a significant $\gamma$-ray signal, the resulting upper limits on the photon flux can still be used to investigate models based on IC emission from leptons accelerated by msPSRs. 
All statements below rely on the assumption that the size of the potential emission region is covered by either of the two analysis methods we employed for this study. 
Given the angular resolution of \hess , a point-like source would basically mean that the leptons are  contained within a $<$4$^\prime$ radius around the GC center.
For the analyses searching for extended emission we assumed that the physical size of the emission region is the same among all GCs, using the intrinsic extent of \htfive\ as a template. 
This would be justified if \htfive\ is indeed related to the msPSR/IC scenario, if the leptons can propagate out of the half-mass region (for some smaller GCs even out of the tidal radius), and if the diffusion and energy-loss mechanisms as well as the age of the lepton population are comparable among all GCs. 

Now, assuming that \htfive\ is indeed related to \terzan\ and that the msPSR/IC scenario is the dominant process for VHE $\gamma$-ray emission, the IC flux should scale as: 
\begin{equation}
\label{eq-IC-flux}
        F_\mathrm{IC} \propto N_\mathrm{msPSR} \times N_\mathrm{ph} \times d^{-2},
\end{equation}
with the number of msPSRs $N_\mathrm{msPSR}$, the total number of available target photons in the volume occupied by relativistic leptons for the IC process $N_\mathrm{ph}$ and the distance to the GC $d$ \citep[Eq.~\ref{eq-IC-flux} is essentially the same as Eq.~8 in][]{2008AIPC.1085..277V}. 
This simple scaling relation is of course only valid if the mechanisms for lepton production are similar among all GCs and, particularly, if the lepton energy spectra and the acceleration efficiencies of the msPSR populations are comparable. 
In this case the proportionality constant in Eq.~\ref{eq-IC-flux} can be fixed using the flux of \htfive , which then allows to calculate the expected VHE $\gamma$-ray flux for the other GCs. 
In addition to the parameters given in Eq.~\ref{eq-IC-flux} also the energy density of the ambient magnetic field in the emission region influences the IC flux due to the competing synchrotron radiation losses. 
However, this only changes our predicted IC fluxes if the magnetic field strengths ($B_\mathrm{GC}$) for the 15 GCs are different from the one in \terzan\ ($B_\mathrm{T5}$). 
In this case the expected IC flux should be scaled by an additional factor of $B_\mathrm{T5}^2/B_\mathrm{GC}^2$. 

The total number of msPSRs is difficult to access directly in a homogeneous way for all 15 GCs in this study. 
This is mainly due to the fact that the various radio surveys feature varying or unknown degrees of completeness, which becomes a very problematic issue for such a large sample of GCs. 
As an alternative to the number count of detected radio msPSRs the total HE $\gamma$-rays flux measured with \fermi /LAT could also be used as a direct indicator for the msPSR population.
However, the GeV measurements suffer from large statistical uncertainties and are only available for two GCs in this sample. 

Here, we rely on a more indirect approach to estimate $N_\mathrm{msPSR}$ by using the stellar encounter rate $\Gamma_\mathrm{e} = \rho_0^{1.5} \times r_\mathrm{c}^2$ which is strongly correlated to the number of neutron star X-ray binaries, the progenitors of msPSRs \citep[see, e.g.][and references therein]{2006ApJ...646L.143P}.
Here $\rho_0$ and $r_\mathrm{c}$ denote the central luminosity density and the core radius of the GC, respectively. 
It has also been shown that $\Gamma_\mathrm{e}$ is correlated with the HE $\gamma$-ray flux \citep{2010A&A...524A..75A,2010ApJ...714.1149H}. 
However, there might be systematic uncertainties in using $\Gamma_\mathrm{e}$ as a proxy for $N_\mathrm{msPSR}$ because the efficiency for the formation of msPSRs also depends on other parameters such as the metallicity of the GC \citep[see e.g.][]{2008MNRAS.386..553I}. 
Assuming that $\Gamma_\mathrm{e}$ is indeed proportional to $N_\mathrm{msPSR}$ and that the proportionality constant is the same for all GCs, Eq.~\ref{eq-IC-flux} then reads: 
\begin{equation}
\label{eq-IC-flux-new}
        F_\mathrm{IC} \propto \rho_0^{1.5} \times r_\mathrm{c}^2 \times N_\mathrm{ph} \times d^{-2}.
\end{equation}

To calculate $F_\mathrm{IC}$ we used the values for $\rho_0$, $d$ and $r_\mathrm{c}$ (converted to units of pc using $d$) as given by H10. 
For $N_\mathrm{ph}$ we used four different target photon fields: 1) the optical light from the GC stars, 2) the Galactic infrared and 3) the Galactic optical background photon fields \citep[as used in GALPROP\footnote{http://galprop.stanford.edu/},][]{2005ICRC....4...77P} at the location of the GC (as given by H10), and finally 4) the cosmic microwave background. 
The number of photons in case 1) is approximately proportional to the V-band luminosity of the globular cluster, which we chose for $N_\mathrm{ph}$ using the values given in H10. 
For cases 2), 3) and 4) $N_\mathrm{ph}$ is proportional to the energy density of the respective radiation fields. 
To compare the expected IC flux with the upper limits we first calculated for each GC the VHE $\gamma$-ray flux of \htfive\ above the respective energy threshold, using the published results \citep{2011A&A...531L..18H}: flux normalization at 1\,TeV: (5.2$\pm$1.1)$\times 10^{-13}$\,cm$^{-2}$\,s$^{-1}$\,TeV$^{-1}$, and power-law spectral index:  $-2.5\pm 0.3_\mathrm{stat}\pm 0.2_\mathrm{sys}$. 
We then used this flux to determine the proportionality constant in Eq.~\ref{eq-IC-flux-new}, which in turn yields the expected IC flux for the respective GC. 
Here we considered two extreme cases, namely that the leptons are either confined within the core region where the GC starlight is the dominant target photon field, or that the particles have propagated far away from the core where the Galactic background fields and the cosmic microwave background become the most important components. 
In the latter case we used the sum of the energy densities of the three photon fields as a proxy for $N_\mathrm{ph}$.
The ratio between the flux upper limits and the expected IC flux are given for both cases in the last two columns of Tab.~\ref{tab-analysis-results}. 

In the simple scaling approach (Eq.~\ref{eq-IC-flux-new}) to estimate the IC flux we assumed that the lepton spectral shapes as well as the acceleration efficiencies of the msPSR populations are comparable among all GCs, which might not be the case. 
Potential deviations could arise from different energy-loss and diffusion timescales, differences in particle (re-)acceleration mechanisms or the spectral shapes of the injection spectra. 
Those and other issues would need to be addressed by more detailed models for each individual GC. 
Apart from the systematics mentioned above, the parameters that are directly used in Eq.~\ref{eq-IC-flux-new} are also subject to uncertainties, which are discussed below. 
The most impact by far for potential uncertainties comes from the distance $d$ because $F_\mathrm{IC}$ itself depends inverse-quadratically on $d$; also the values of $\rho_0$ as well as $r_\mathrm{c}$ are less strongly dependent on $d$. 
Unfortunately, the uncertainties of $d$ are perhaps the most difficult to quantify because there are several different methods for distance determinations of GCs, each suffering from other systematic issues. 
Therefore, we decided here to use the distances quoted by H10 for all of the GCs studied in this work, to only rely on one method. 
Assuming an error margin for $d$ of a factor of two (based on typical variations between different distance estimation methods) yields an uncertainty of a about a factor of five in the expected IC flux. 
Additionally, the confidence intervals of the spectral parameters of \htfive\ \citep{2011A&A...531L..18H} introduce further uncertainties in the expected IC flux of about 20\,\%, yielding a total error margin of a factor of $\sim$5. 
To estimate the uncertainties of the predicted model flux from the full stacks, we assumed a factor of 5 uncertainty for each individual GC and propagated the errors. 
We show the resulting uncertainty band for the stacking analyses in the last two columns of Tab.~\ref{tab-analysis-results}. 

With the optical light of the GC member stars as the target photon field, the flux upper limits for point-like emission from the two GCs NGC\,6388 and NGC\,7078 lie about one order of magnitude below the expected IC flux (see Tab.~\ref{tab-analysis-results}). 
This is mainly due to the fact that \terzan\ is relatively faint in the V-band compared to many other GCs, which yields large expected IC fluxes for GCs with a similar or even larger stellar encounter rate $\Gamma_\mathrm{e}$. 
Among all GCs studied here, given the uncertainties related to the predictions, only in these two cases the simple scaling model is challenged by our observational results. 
However, the fact that \htfive\ is extended and offset beyond the half-mass-radius of \terzan\ could point towards fast diffusion and long energy-loss timescales of the lepton population in \terzan . 
If this is indeed the case, then other target photon fields would be dominant over the GC starlight. 
Therefore, the results of the point-like analyses are better suited to be compared to the predictions with the GC starlight as target photon fields, whereas the results from the extended analyses are more relevant for the predictions based on Galactic background photons. 
We note here that for some GCs the Galactic background could still make a contribution to the IC flux even for regions close to the GC cores, an issue that is not taken into account in our scaling model. 
Looking at the results for the infrared and optical background photon fields, the upper limits do not allow for similarly strong constraints as in some cases when considering the GC starlight. 
Given the uncertainties related to the parameters in Eq.~\ref{eq-IC-flux-new}, the msPSR/IC scaling model cannot be ruled out for any of the GCs assuming Galactic background photons as IC target. 

To make a statement about the GC sample as a whole, we derived the expected IC flux from the total stack by calculating the weighted mean flux from all GCs. 
As weights we used the acceptance-corrected livetimes of the individual clusters. 
The upper limits for the stack are about two orders of magnitude below the expectation for the GC starlight and at a level of $\sim$50\% for the external target photon fields (see bottom section of Tab.~\ref{tab-analysis-results}). 
In the first case, even considering the uncertainties, this seems to be quite a strong constraint on the simple msPSR/IC scaling model. 
Therefore, either the IC flux from GCs does not scale as expected, or the VHE $\gamma$-ray source near \terzan\ is not related to msPSRs. 
However, for the second case, i.e. assuming background photon fields as the dominant IC target, the upper limits for the stack are compatible with expected fluxes, and therefore do not allow to put similarly strong constraints on the model. 

There are many possibilities why \terzan\ may stand out among the general GC population. 
In the msPSR/IC model the confinement of leptons in the emission region might be stronger in \terzan\ compared to other GCs, which would lead to an accumulation of particles from earlier epochs that would therefore enhance the observed IC flux per msPSR. 
However, because \htfive\ is extended and offset from the core, the trapping mechanism is required to work on similarly large spatial scales. 
A high interstellar magnetic field strength in the vicinity of \terzan\ could provide such a trapping mechanism, but would also increase the synchrotron losses and therefore decrease the expected IC flux. 
These competing effects would need to be modeled in detail for each individual GC, and the results presented in this paper might yield interesting constraints. 

As already mentioned in sect.~\ref{sec-introduction} there are also models that are not related to the msPSR/IC scenario which suggest that the VHE $\gamma$-ray emission is due to hadronic processes in a short gamma-ray burst remnant \citep{2011A&A...533L...5D}. 
Due to the extremely low rates of these catastrophic events of only $\sim$10$^{-4}$\,yr$^{-1}$ \citep{2007PhR...442..166N} it could be easily explained why only one GC stands out among the whole population. 

\section{Conclusions}
\label{sec-conclusions}
Judging from recent HE and VHE $\gamma$-ray observations, GCs could feature interesting high-energy processes,   
and, in particular, could constitute a new class of sources in the VHE $\gamma$-ray regime, if the recently detected source \htfive\ is indeed related to the GC \terzan . 

To test the validity of the msPSR/IC scenario for VHE $\gamma$-ray emission, we analyzed the \hess\ data of 15 Galactic GCs. 
According to our observations, neither any individual GC nor the stacking analyses show significant excess emission above the estimated background. 
Based on a simple model for IC emission and the measured VHE $\gamma$-ray flux of \htfive\ we calculated predictions for the fluxes for each individual GC and compared them with the derived upper limits. 

The non-detections of the two GCs NGC\,6388 and NGC\,7078 challenge the simple IC scaling model when assuming the GC starlight as the dominant target photon field. 
However, when searching for VHE $\gamma$-ray emission from a larger scale, comparable to what is seen from \terzan , and considering Galactic optical, infrared and CMB photons as the dominant IC target, no clear constraints can be put on our simple msPSR/IC scaling model. 
The upper limits derived from the stacking analyses are about two orders of magnitude below the expected cumulative IC flux from the whole GC sample when assuming the GC starlight as the dominant target photon field. 
Taking the uncertainties of those parameters into account that directly influence our flux predictions, these results still exclude the simple msPSR/IC scaling model for this case. 
But, as mentioned in sect.~\ref{sec-discussion}, there are several caveats to the simple scaling model applied here that would have to be considered in more elaborate and custom-tailored models for each individual GC. 
Also, there is still the possibility that the VHE $\gamma$-ray signal from \terzan\ might not be related to the msPSRs at all and arises from hadronic or other processes. 

Further studies of GCs are still ongoing and future deep observations with Cherenkov telescopes and X-ray observatories of the most promising candidates will increase the strength of the derived upper limits, or might even facilitate another detection next to \terzan .

\begin{acknowledgements}
The support of the Namibian authorities and of the University of Namibia
in facilitating the construction and operation of H.E.S.S. is gratefully
acknowledged, as is the support by the German Ministry for Education and
Research (BMBF), the Max Planck Society, the German Research Foundation (DFG), 
the French Ministry for Research,
the CNRS-IN2P3 and the Astroparticle Interdisciplinary Programme of the
CNRS, the U.K. Science and Technology Facilities Council (STFC),
the IPNP of the Charles University, the Czech Science Foundation, the Polish 
Ministry of Science and  Higher Education, the South African Department of
Science and Technology and National Research Foundation, and by the
University of Namibia. We appreciate the excellent work of the technical
support staff in Berlin, Durham, Hamburg, Heidelberg, Palaiseau, Paris,
Saclay, and in Namibia in the construction and operation of the
equipment.
\end{acknowledgements}

\bibliographystyle{aa}
\bibliography{citations_v2}

\end{document}